\documentstyle[aps,prl,epsf]{revtex}
\begin{document}

\def\wig#1{\mathrel{\hbox{\hbox to 0pt{%
    \lower.5ex\hbox{$\sim$}\hss}\raise.4ex\hbox{$#1$}}}}
\def\lsim{\wig <}
\def\gsim{\wig >}

\draft
\tighten
\twocolumn[\hsize\textwidth\columnwidth\hsize\csname @twocolumnfalse\endcsname

\title {Patterns of Striped Order in the
        Classical Lattice Coulomb Gas }
\author{Sung Jong Lee$^1$, Jong-Rim Lee$^2$, and Bongsoo Kim$^3$}
\address{ $^{1}$ Department of Physics,   
The University of Suwon, Hwasung-Gun, Kyunggi-Do 445-743, Korea }

\address{ $^{2}$ Division of Electronics, Computer and Telecom.
Engineering, Pukyong National University, Pusan, 608-737, Korea }

\address{ $^{3}$ Department of Physics,   
 Changwon National University, Changwon 641-773, Korea }

%\today

\medskip

\maketitle
\begin{abstract}

       We obtain via Monte Carlo simulations the low temperature  charge
       configurations in the lattice Coulomb gas on square lattices for
       charge filling ratio $f$ in the range $1/3 < f < 1/2 $. 
       We find a simple regularity in the low temperature charge configurations 
       which consist of a suitable periodic combination of a few basic striped 
       patterns characterized by the existence of partially filled diagonal 
       channels. In general there exist two separate transitions where the lower  
       temperature transition ($T_p$) corresponds to the freezing of charges 
       within the partially filled channels. $T_p$ is found to be sensitively
       dependent on $f$ through the charge number density $\nu = p_{1}/q_{1}$
       within the channels.

\end{abstract}


\vskip2pc] \narrowtext

\pagebreak

Competition between periodic pinning potential and repulsive inter-particle
interactions often generate interesting frustration effects on the equilibrium
properties of particle systems resulting in diverse commensurate or incommensurate
phases \cite{hubbard,pokrovsky,pbak}. 
In the simplest case of one dimensional system on a lattice, there exists 
a well known prescription of obtaining the ground state configuration  \cite{hubbard,pokrovsky} 
for a given particle density provided that the inter-particle potential is convex and vanishing
at infinity. In two dimensions, systematic or analytic results are scarce \cite{watson}.  

 One of the extensively studied model systems in two dimensions is the classical lattice
Coulomb gas (LCG), where a fraction 
$f=p/q$ ($p$ and $q$ are relative primes) of the lattice 
sites are occupied by unit charges upon uniform neutralizing background \cite{cg-jja}. 
LCG appears in the vortex representation of the two dimensional Josephson junction
arrays under uniform external magnetic field via Villain approximation \cite{villain}
of the uniformly frustrated XY (UFXY) model, where the parameter $f$ 
corresponds to the ratio of the magnetic flux per unit plaquette and the 
superconducting  flux quantum $\Phi_0 = hc/2e $.    
In these systems, physical properties such as transition 
temperatures or critical currents can depend sensitively on (the rationality of) 
the value of the frustration parameter $f$ due to the commensuration effect \cite{rational}.

In spite of many years of research effort, there still remains wide 
range of $f$ parameter (notably for $1/3 < f < 1/2$) for which not much is known 
in terms of equilibrium properties. Even the ground state 
configurations are known only for some limited cases of values of $f$, 
especially for low order rationals \cite{teitel}.
In this paper, we report simulation results on the low temperature ordered 
structures  of LCG on square lattices for $1/3 < f  < 1/2 $.

In the case of UFXY model on square lattices, Halsey proposed
staircase states \cite{staircase} as the low temperature ordered configurations 
with periodicity $q \times q$. However, the staircase state turns out to be the true ground 
state configuration only for some limited values of $f$ such as $f=1/2$, $1/3$, $2/5$, $3/8$,
etc.  More recent works on UFXY model \cite{dt} and LCG \cite{gtg} shows that for $f=p/q$
near $1-g$ ($g= (\sqrt{5}-1)/2 $) with $q$ large, the low temperature vortex configuration
in the UFXY model can be different from the ordered charge configurations in LCG.
Especially, in the case of LCG on square lattices, 
it was shown \cite{gtg} that there appear diagonal striped configurations with 
partially filled diagonals. However, the exact patterns of low temperature charge 
configurations for general $f$ of dense charge filling have not been enumerated yet. 

In this paper, we find numerically for LCG on square lattices with $f$ in 
a large part of the range $1/3 < f < 1/2$, that, below a first order 
transition temperature $T_c$, the charge configuration consists of periodic
arrangements of a few basic striped patterns. Four regimes of values of $f$ 
are identified, each of which represents a specific class of striped patterns 
characterized by the existence of partially filled diagonals.   
$T_c$ is found to be only weakly dependent on $f$. 
Below $T_c$, there exist in general a temperature range where 
the charges within partially filled diagonal channels are disordered (thus mobile
along channels).    
We found that there exist another transition at a lower temperature $T_{p}$ at which
freezing of charges within partially filled channels occur. An interesting result is
that this lower transition temperature depends sensitively on the rationality 
of the charge number density within channels $\nu = p_{1} / q_{1}$, decreasing  
monotonically in $q_{1}$.

General 2D LCG \cite{jrl} is described by the following Hamiltonian,

\begin{equation}
{\cal H}_{CG}={1 \over 2}\sum_{ij}Q_iG(r_{ij})Q_j \label{eq:Hcg}
\end{equation}
where $r_{ij}$ is the distance between the sites $i$ and $j$,
and  the magnitude of charge $Q_i$ at site $i$ can take either $1-f$ or $-f$.
Charge neutrality condition $\sum_iQ_i=0$ implies that the number density of 
the positive charges is equal to $f$. 
We can thus view the system as a lattice gas 
of $N \cdot f$ charges of unit magnitude upon uniform negative 
background charges of charge density $-f$ ($N= L^2$ is the total size
of the system with the linear dimension $L$).  
The lattice Green's function $G(r_{ij})$  solves the equation
$(\Delta^2 -   \lambda^{-2}) G(r_{ij})=-2\pi\delta_{r_{ij},0}$,
where $\Delta^2$ is the discrete lattice Laplacian and $\lambda$ is 
the screening length. For the case of usual Villain transformation of UFXY model,
we have $\lambda = \infty$. But the screening term is included in this equation
for generality. 
On a square lattice with periodic boundary conditions, $G(\vec{r})$ is given by 

\begin{equation}
G(\vec{r})={\pi \over N} \sum_{\vec k \neq 0} { e^{i\bf{k} \cdot r}-1 \over 
{2 - \cos k_x -\cos k_y + 1/\lambda^2} } ,
\end{equation}
where $\bf{k}$ are the allowed wave vectors with $k_{\mu} = 2 \pi n_{\mu}/ L$ 
($\mu=x, y$), with $n_{\mu} = 0, 1, \dots, L-1 $.
In the case of infinite screening length, for large separation $r$,
 one gets $G(\vec r)\simeq -\ln r$ \cite{jrl}. In this work, the presented results 
are all obtained for the case of $ \lambda \rightarrow \infty $  \cite{screen}. 

In our MC simulations, the initial disordered random configuration 
is updated according to the standard Metropolis algorithm
by selecting a positive charge at random and moving it over to 
one of the {\em nearest neighbor (NN)} or {\em next nearest neighbor (NNN)} 
sites \cite{gtg}.  
 An important aspect of our simulations is that one has to choose 
the lattice size appropriately in order to match the periodicity of the 
low temperature configuration (see below). If, otherwise, 
one chooses a lattice size that is incommensurate with the periodicity 
of striped patterns, then one ends up with defective charge configurations.

To begin with, let us present the four basic component patterns (Fig.~1) before
going into the full description of the striped configurations. 
First component pattern (I) is a sequence of three diagonals 
which are {\em empty, filled}, and {\em empty} respectively (that may be 
denoted by {\bf (010)} in our notation where {\bf 1} refers to a filled 
diagonal and {\bf 0} refers to an empty diagonal). In other words, it is 
a pattern with single isolated diagonal filled with charges, that is neighbored 
by empty diagonals on both sides. Repetition of this pattern produces the
ground state configuration for $f=1/3$ with spatial periodicity three.  

Second component pattern (II) consists of a sequence of five 
diagonals that can be written as {\bf (01010)}. This forms the basis
of the ground state configuration for $f=2/5$ with 
lattice periodicity five.   
The third component pattern (III) consists of a sequence
of seven diagonals that can be denoted by {\bf (010p010)} 
where {\bf p} refers to a partially filled diagonal where only part 
of the diagonal sites are occupied by positive charges. This is 
essentially a partially filled diagonal enveloped by one filled diagonal 
on both sides at second neighbor 
diagonal position, which may be termed as a {\em channel} structure. 
This can form a basis of a periodic configuration with spatial lattice
periodicity seven.
Lastly, the fourth component pattern (IV) can be denoted by
{\bf (01010p01010)}. This is component pattern consisting of a partially
filled diagonal bounded by two type II stripe patterns on both sides.
This can form a basis of a periodic configuration with spatial lattice
 periodicity eleven.

 We are now in a position to be able to give a detailed description of 
the low temperature charge patterns. 
We may identify four regimes for the striped charge configuration.
We may call these by regime $A$, $B$, $C$, and $D$ respectively.
Fig.~2 shows typical representative configuration in each of the four regimes.

In regime $A$ that is bounded by $1/3 \lsim f \lsim f_{c1}$ ($f_{c1} \simeq 0.357$),
the low temperature charge configuration consists of combinations 
of type I and III patterns, where $l$ ( $l=1,2,3 \cdots$ ) copies
of type I patterns in sequence followed by a single 
type III pattern ($\rm{I}^{\it{l}} \rm{III}$) forms a basic unit, repetition of which 
forms the whole charge configuration. We can easily see that the lattice 
periodicity of the ordered stripe configuration is equal to $p_{A} = 3l+7$. 
As $l$ approaches infinity, we recover the case of $f~=1/3$. 
As the value of $f$ increases within regime $A$, the value of $l$ decreases
monotonically in step-like manner. Therefore, regime $A$ is further divided
into sub-regimes each of which is characterized by a positive integer $l$.

Regime $B$ covers the region $f_{c1} \lsim f \lsim f_{c2}$ ($f_{c2} \simeq 0.381$),
where the ordered charge configuration simply 
consists of repetitions of type III stripe patterns with the resulting 
lattice periodicity $p_{B} = 7$.
In regime $C$ which is bounded by $f_{c2} < f < 2/5$,
the low temperature charge configuration consists of combinations of type II and 
III patterns, with $m$ ($m=1,2,3 \cdots$) copies of type II stripe patterns 
in sequence followed by a single type III pattern ($\rm{II}^{\it{m}} \rm{III}$) forms
a basic unit. 
Here, we can see that the lattice periodicity of the ordered stripe
configuration is equal to $p_{C} = 5m+7 $. As the value of $f$
increases within regime $C$, the value of $m$
increases monotonically in step-like manner. The case of $f~=2/5$
corresponds to the limit of $m \rightarrow \infty$.

Regime $D$ corresponds to $2/5 < f \lsim f_{c3}$ ($f_{c3} \simeq 0.425$), 
where the unit period of
ordered configuration consists of combinations of a type IV pattern plus
 $n$ ($n=0,1,2,3, \cdots$) repetitions of type II stripe patterns that may
 be denoted by $ \rm{II}^{\it{n}} \rm{IV}$. 
 We see that the lattice periodicity of the ordered stripe
configuration in regime IV is equal to $p_{D} = 5n+11$. As the value of $f$
increases within regime $D$, the value of $n$
decreases monotonically in step-like manner.

Note that the periodicity in the above refers 
to the periodicity of the filled diagonals only, neglecting the true 
periodicity including the charge configurations within the partially 
filled diagonals.
The true spatial periodicity of the charge configurations can be many times
larger than the stripe periodicity since we also have to take into account
the correlation of charge configurations between different partially filled
channels.      

For any given value of $f$, we can easily obtain the filling density $\nu$ inside 
the partially filled channel using the relations $f = (l+2 + \nu) / (3l+7)$
(for regime $A$ and $B$),  $f= (2m+2 + \nu) / (5m+7)$ (regime $C$) and
$f=(2n+4 + \nu)/(5n+11)$ (regime $D$) respectively where $l$, $m$, $n$ and $\nu$ 
are as defined above.
 As the value of $f$ continuously increases within one sub-regime, the system 
in the low temperature stable configuration simply adjusts itself by accomodating 
the extra number of charges into the partially filled diagonal channels and thereby
changing the charge filling $\nu$ within the channels.

We find that there exists in general another transition at lower temperature 
$T_{p}$ corresponding to charge freezing inside partially filled channels.
In order to check that, we calculated the inverse dielectric constant along 
two perpendicular diagonals to identify the transition temperatures. 
The wave-vector dependent inverse dielectric constant is defined as follows \cite{epsilon},

\begin{equation}
\epsilon^{-1}(\vec{k})= \left (  1-{{2 \pi } \over {T \Omega k^2 }}
< \rho_{k} \rho_{-k} > \right ),
\end{equation}
where  $\rho_{k} \equiv \sum_{r_{i}} Q(r_{i}) \exp (-i \vec{k} \cdot \vec{r_{i}})$
is the Fourier component at wave-vector $\vec{k}$ of the charge density 
and $\Omega \equiv L^2 $ is the total area of the system. 
By letting $k \rightarrow 0$ for a given direction of the wave-vector,
one can obtain the long wavelength limit of the inverse dielectric constant along 
a specific direction.

 Figure 3 shows the dependence of the inverse dielectric constant along parallel
and perpendicular to the stripes respectively for $f=13/35$.
We can clearly see that there exists, in addition to the transition at
$T_c \simeq 0.32$ corresponding to the onset of striped order, intermediate 
regime of temperature where the dielectric constant exhibits asymmetry due to 
the channel-striped structures. Also, the system is seen to undergo another
transition at lower temperature $T_{p} \simeq 0.014$ (determined rather 
arbitrarily as the temperature where the inverse dielectric constant is 
equal to $0.4$). 
Figure 4 shows the dependence of the inverse dielectric constant along parallel
to the stripes for various values of $f$ in regime $B$, where a wide variation 
is seen in 
the temperature dependence of the inverse dielectric constant parallel 
to the channels.

Shown in Fig.~5 is the dependence of the two transition temperatures on $f=p/q$
in regime $B$ where the higher transition $T_c$ is seen to depend on the values
of $f$ smoothly, while the lower transition temperature $T_p$ exhibits sensitive
dependence on $f$. From the inset of Fig.~5 where a plot of $T_{p}$ versus the integer 
denominator $q_{1}$ of the charge number density $\nu \equiv p_1 /q_1$ within channels, 
one can recognize that $T_{p}$ decreases monotonically as $q_{1}$ increases. 
This is another commensuration effect coming from the rationality of the 
particle number density within channels each of which forms effectively 
a one-dimensional lattice gas system if we neglect the interaction between 
different channels. 
The arrangement of charges within each channel was found to follow the pattern
given in ref. \cite{hubbard} and \cite{pokrovsky} at least for the
values of $f$ considered in this work.

 One can ask what determines the boundary value of $f$ and 
$\nu$ for each of the sub-regimes, in other words, what is the stability 
criterion for each pattern of the striped configuration. Even though an 
analytic formula cannot be given, our simulations suggest that the criterion 
of determining the crossover point between two neighboring subregimes is 
related to the electrostatic stability which is determined by the filling 
density inside the partially filled channel. We found numerically that 
for any given value of $f$ the filling density $\nu$ in the channel 
approximately satisfies the inequality $0.4 \lsim  \nu \lsim 0.7$. 
For a given sub-regime, possible values of $\nu$ were always within this bound.
Beyond some threshold value $\nu_c$ of $\nu$ that is within the above bound, 
electrostatic instability begins to set in, and rearrangement of the whole 
charge configuration occurs in order to form a new stable ordered patterns. 

Now we are left with the region of $f_{c3} < f < 1/2$.
Even though we have not investigated the ordered configurations extensively
for all values of $f$ in this regime, we could see, from some of our annealing
simulations for rational values of $f$ in this regime, that the low temperature  
configuration no longer shows striped patterns but rather 
consists of regular arrays of hole defects upon $f=1/2$ checkerboard 
configurations. 
If we suppose that $f \equiv 1/2-\alpha \equiv (1-p_2 / q_2 )/2$
with $p_2$ and $q_2$ relative primes, then we can easily see that
$f' \equiv p_2 /q_2 \equiv 2\alpha $ represents the density of 
 hole defects upon the $f=1/2$ checkerboard configuration.
Therefore, we can suspect that the low temperature
defect configuration in this regime will be equivalent to the charge configuration
of lattice Coulomb gas with $f=f'$. We could confirm this expectation for a few cases
of $f$ \cite{teitel,defect}.

 In summary, we have shown numerically that the 2D LCG on a square lattice
exhibits a simple regularity in its striped charge configuration 
at low temperatures for filling factor $f$
in a large part of the range $1/3 < f < 1/2$ which is characterized by the
existence of partially filled diagonals.
The low temperature ordered configuration consists of a 
simple combinations of four basic striped patterns. In general, there exists 
another transition at a lower temperature $T_p$ corresponding to the 
freezing within partially filled channels.  
It would be interesting to observe these  striped charge patterns experimentally,
{\it e.g.}, in regular square arrays of ultrasmall tunnel junctions.
It may also be possible to observe similar patterns
in the macroscopic systems of dielectric charged spheres \cite{granular} under 
periodic pinning potentials.     

This work was supported by the Korea Research Foundation Grant
(KRF-1999-015-DP0098) (SJL, BK) and (KRF-1998-15-D00089) (JRL).

\centerline {\bf FIGURE CAPTIONS}

\renewcommand{\theenumi}{Fig.~1}
\begin{enumerate}
\item
Regimes of charge patterns for the range of value of $f$ between
$1/3$ and $0.425$. Filled squares and empty squares represent positive
($Q=1-f$) and negative ($Q=-f$) charges respectively, while gray squares
denote lattice sites forming partially filled diagonal channels,
where only finite fraction $\nu$ of the sites are filled with positive 
charges.

\end{enumerate}

\renewcommand{\theenumi}{Fig.~2}
\begin{enumerate}
\item
Low temperature charge configurations for (a) regime $A$ with $f=7/20$ 
($l=1$), (b) regime $B$ with $f=13/35$,
(c) regime $C$ with $f=0.384$  ($m=1$) and (d) regime $D$ with $f=0.41$ ($n=1$),
respectively. 
\end{enumerate}

\renewcommand{\theenumi}{Fig.~3}
\begin{enumerate}
\item
Inverse dielectric constants along parallel and perpendicular to the 
stripes for $f=13/35$ and $L=35$ versus temperature. We can clearly see 
an anisotropy of the inverse dielectric constant. 
\end{enumerate}

\renewcommand{\theenumi}{Fig.~4}
\begin{enumerate}
\item
Inverse dielectric constant along parallel to the 
stripes for various vaues of charge number density $\nu$ within channels. All of
the systems shown are chosen from regime $B$.  
\end{enumerate}

\renewcommand{\theenumi}{Fig.~5}
\begin{enumerate}
\item
 Transition temperatures (both $T_c$ and $T_p$) versus the charge filling ratio $f$.
Inset shows the dependence of $T_p$ on the denominator
$q_{1}$ of $\nu = {p_{1} / q_{1}}$.  
\end{enumerate}


\begin{thebibliography}{99}
\bibitem{screen}
 For finite screening length $\lambda$, we find that there exist a
threshold magnitude $\lambda = \lambda_c$ ($\lambda_c \simeq 1.6$ for $f=13/35$)
above which the channel-striped configuration remains stable but
it becomes unstable below $\lambda_c$.
\end{thebibliography}
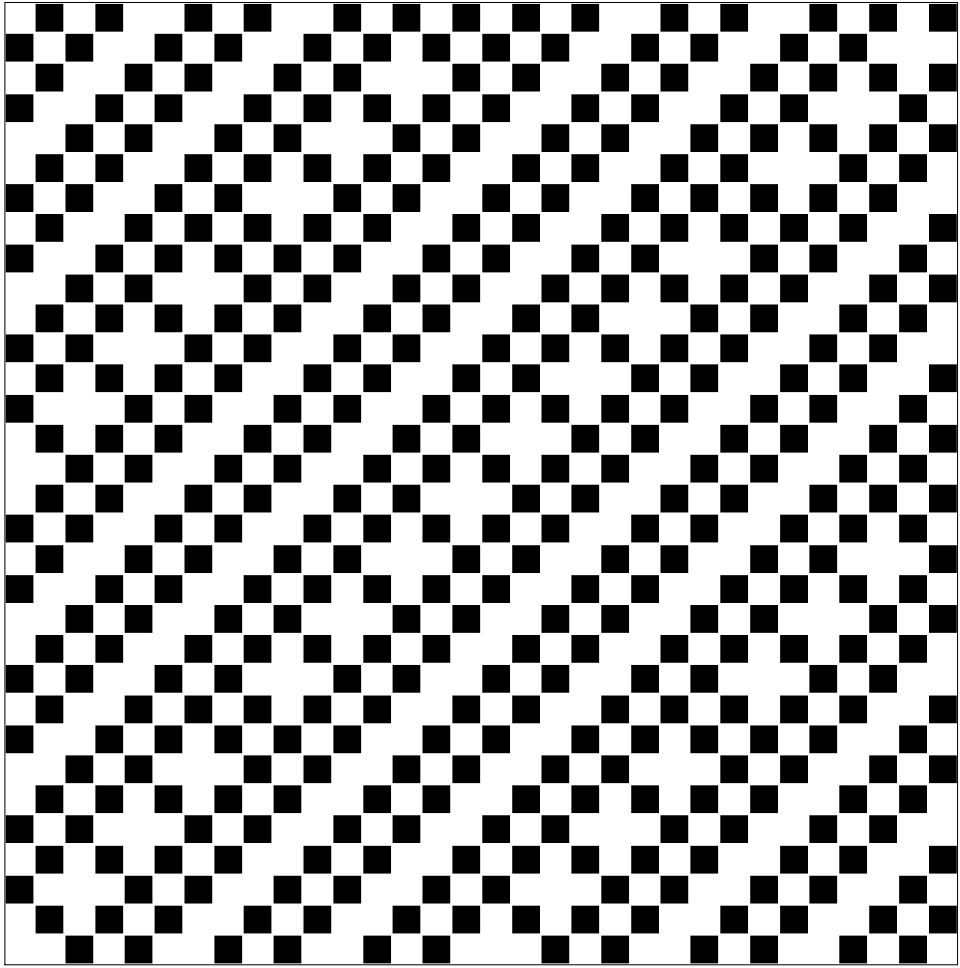
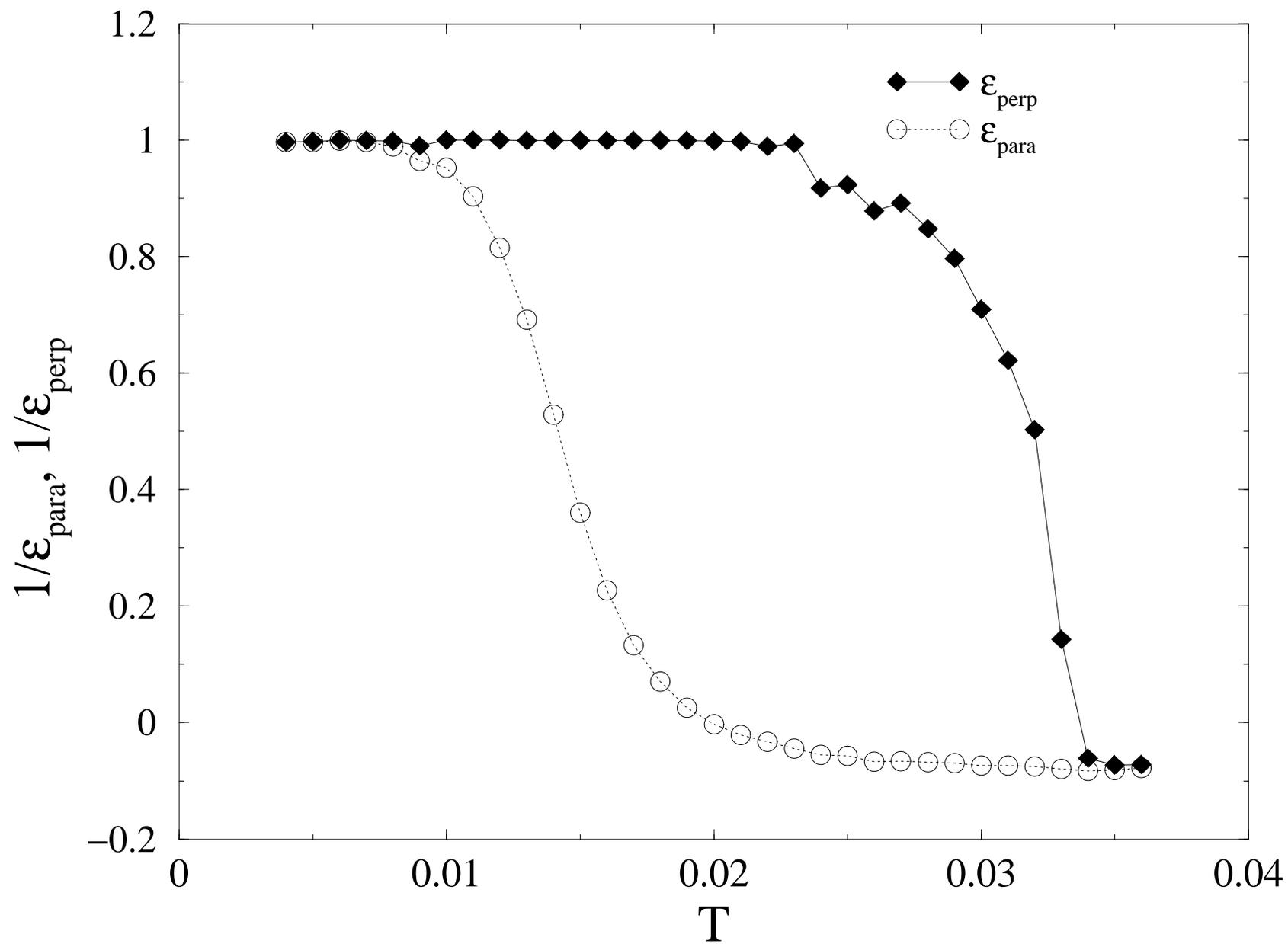
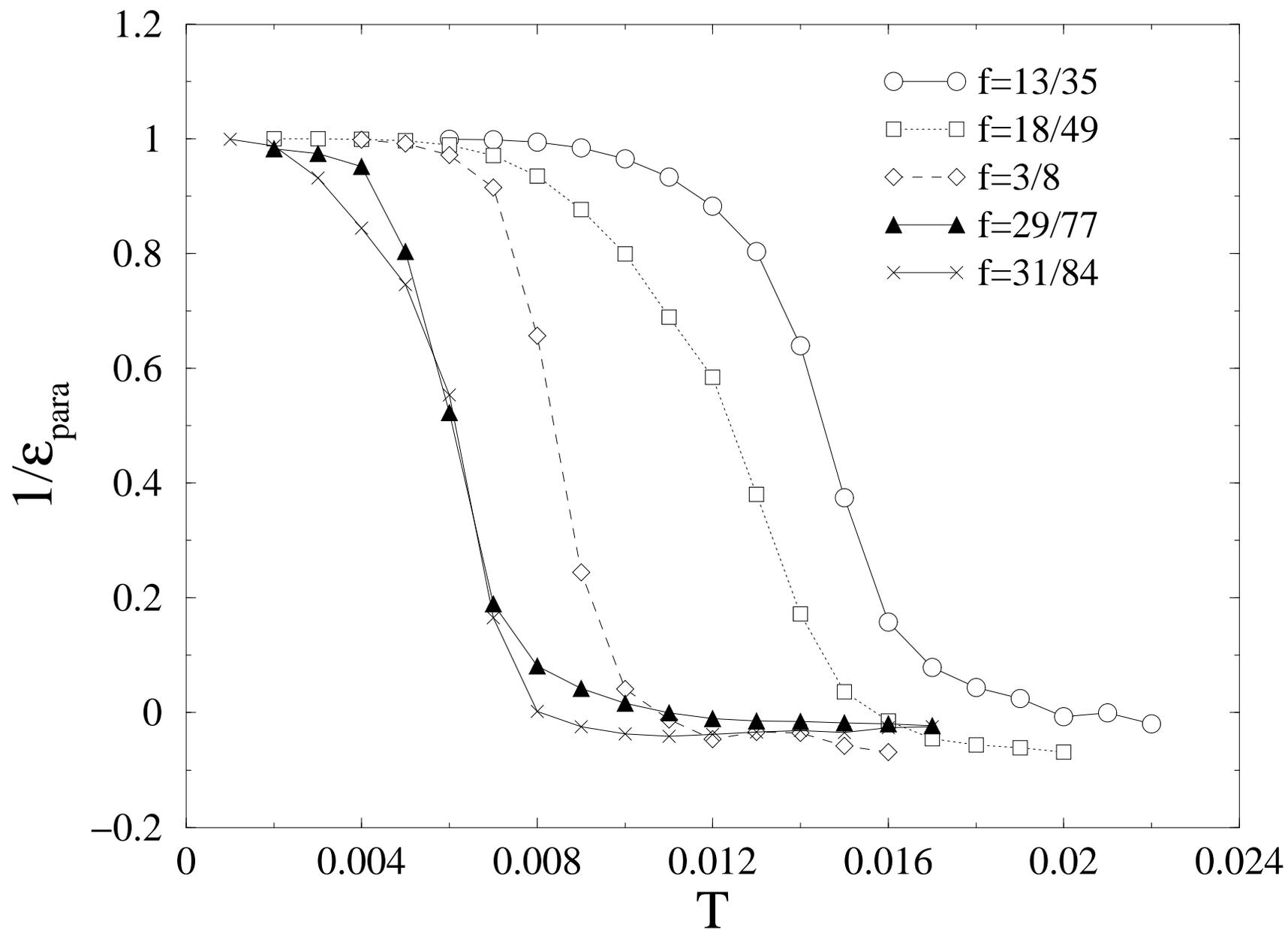
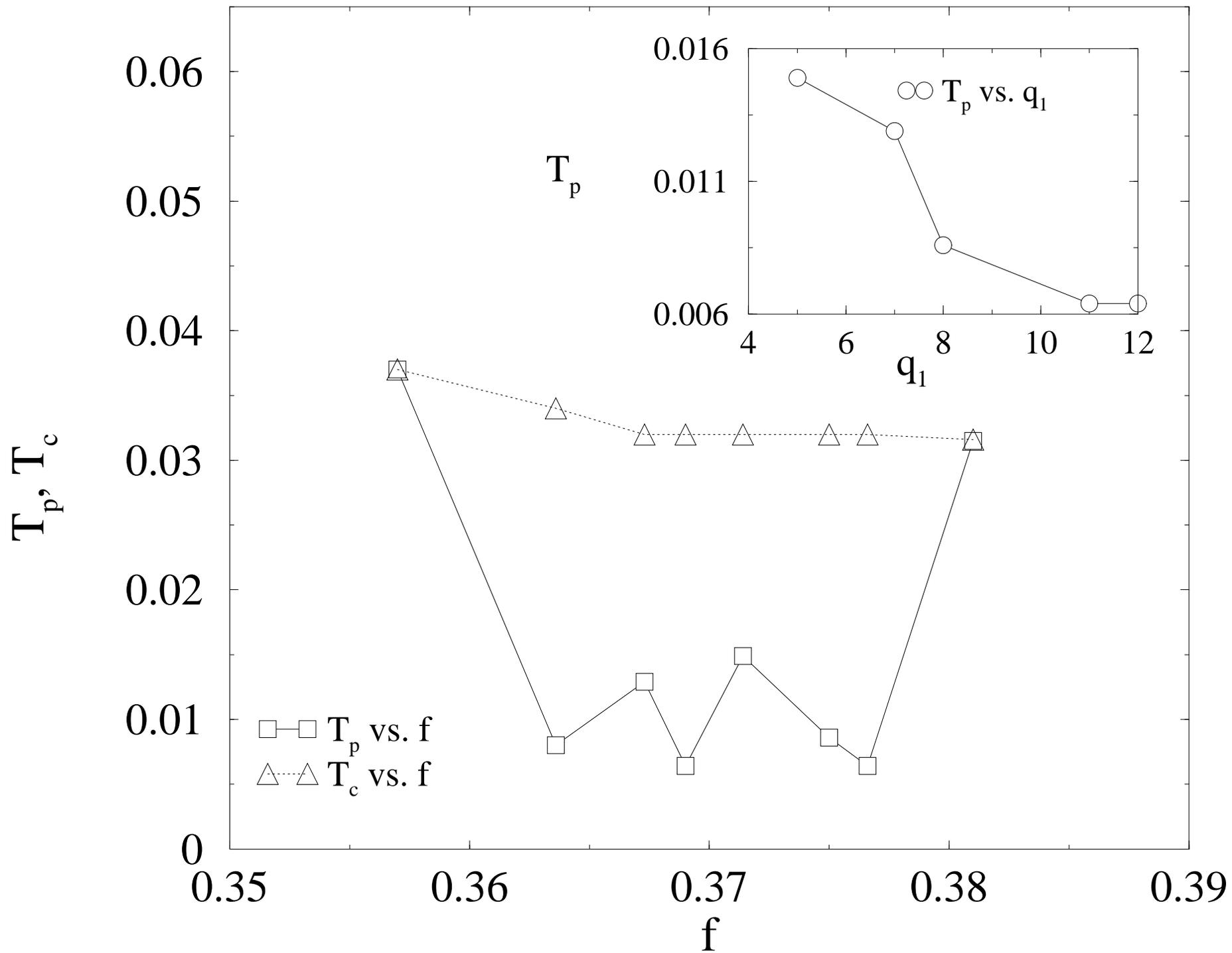
\end{document}